# Single-crystalline Aluminum Nanostructures on Semiconducting GaAs Substrate for Ultraviolet to Near-infrared Plasmonics

Hsuan-Wei Liu,[1] Fan-Cheng Lin,[2] Shi-Wei Lin,[1] Jau-Yang Wu,[1]

Sheng-Di Lin[1] and Jer-Shing Huang [2,3,*]

[1] Department of Electronics Engineering, National Chiao Tung University, Hsinchu 30010, Taiwan

2 Department of Chemistry, National Tsing Hua University, Hsinchu 30013, Taiwan

3. Frontier Research Center on Fundamental and Applied Sciences of Matters, National Tsing Hua University, Hsinchu 30013, Taiwan

AUTHOR ADDRESS: jshuang@mx.nthu.edu.tw

KEYWORDS   Surface plasmon resonance, Ultraviolet plasmonics, Molecular beam epitaxy, Single-crystalline aluminum, Semiconducting substrate, Photoluminescence, Nanoholes, Nanoslits

**ABSTRACT:** Aluminum, as a metallic material for plasmonics, is of great interest because it extends the applications of surface plasmon resonance into the ultraviolet (UV) region and excels noble metals in the natural abundance, cost and compatibility with modern semiconductor fabrication process. Ultrasmooth single-crystalline metallic films are beneficial for the fabrication of high-definition plasmonic nanostructures, especially complex integrated nanocircuits. The absence of surface corrugation and crystal boundaries also guarantees superior optical properties and the applications in nanolasers. Here, we present UV to near-infrared (NIR) plasmonic resonance of single-crystalline aluminum nanoslits and nanoholes. The high-definition nanostructures are fabricated with focused ion-beam (FIB) milling into an ultrasmooth single-crystalline aluminum film grown on a semiconducting GaAs substrate with molecular beam epitaxy (MBE) method. The single-crystalline aluminum film shows improved reflectivity and reduced two-photon photoluminescence (TPPL) due to the ultrasmooth surface. Both linear scattering and non-linear TPPL are studied in detail. The nanoslit arrays show clear Fano-like resonance and the nanoholes are found to support both photonic modes and localized surface plasmonic resonance. We also found that TPPL generation is more efficient when the excitation polarization is parallel rather than perpendicular to the edge of the aluminum film. Such counter-intuitive phenomenon is attributed to the high refractive index of the GaAs substrate. We show that the polarization of TPPL from aluminum well preserves the excitation polarization and is independent of the crystal orientation of the film or substrate. Our study gains insight into the optical property of aluminum nanostructures on high-index semiconducting GaAs substrate and illustrates a practical route to implement plasmonic devices onto semiconductors for future hybrid nanodevices.

Resonant plasmonic nanostructures can concentrate light and enhance sub-wavelength light-matter interaction [1, 2]. In recent years, surface plasmon resonance in nanostructures has been proposed and applied to achieve optical-frequency integrated nanocircuits [3-5], high-performance photovoltaic devices [6], ultrasensitive biochemical sensors [7, 8], enhanced circular dichroism [9], optical trapping [10, 11], efficient steam generation [12], hot electron generation [13] and ultra-small color filters [14]. Silver and gold are typical plasmonic materials for visible to infrared spectral regime. Although silver and gold have relatively good chemical stability and well-known chemistry of synthesis [15, 16], their application in the ultraviolet spectral regime are restricted by the loss due to interband transition [1]. Recently, there are considerable efforts in non-metal plasmonics using materials such as graphene [17, 18], doped semiconductors [19] or nitride materials [20, 21]. However, none of these materials can practically support surface plasmon resonance in the UV regimes. To extend the frequency window of plasmonics into the UV regime, aluminum has been proposed and used [22-25]. From the perspective of light-matter interaction, aluminum plasmonics is useful because it offers nanoscale confinement and enhancement of electromagnetic energy in the UV regime, where electronic transition with energy difference larger than 3.0 eV occurs. In addition, the non-linear optical response of aluminum is also much larger compared to gold and silver, making aluminum a suitable material for higher harmonic generation. From a point of view of industrial production, aluminum is advantageous because it is relatively high natural abundant, low-cost and compatible to the well-developed CMOS and semiconductor manufacturing processes. These advantages make aluminum a promising material for future integration and mass production of plasmonic-semiconducting hybrid nanodevices with ultra-high operational frequency in the UV regime and sub-wavelength footprints at the nanoscale.

So far, aluminum plasmonic nanostructures are mainly prepared by e-beam lithography or focused-ion beam (FIB) milling into thermally evaporated multi-crystalline aluminum film. The random crystal grains and surface roughness of the multi-crystalline aluminum film can directly lead to structural defects and increase the scattering loss of surface plasmons [26, 27]. Compared to gold and silver films, evaporated aluminum films usually show worse quality even at the optimal evaporation condition. The low film quality makes it very difficult to fabricate complex nanocircuits containing multiple circuit elements and greatly limits the application of aluminum, for example, in plasmonic nanolasers, which requires an ultrasmooth surface of metallic film [28]. To improve the film quality, efforts have been made by optimizing the thermal evaporation condition [29, 30]. However, the random orientation and boundaries of crystal grains in the multi-crystalline aluminum film still hamper the fabrication precision and product yield. Moreover, the crystal boundaries and voids in the film can also lead to variation of the optical properties [31], which limits the realization of optimally designed plasmonic nanostructures. For these reasons, using atomically flat single-crystalline aluminum film to fabricate high-definition plasmonic nanostructures is of fundamental and practical importance.

Although single-crystalline gold and silver microplates from chemical synthesis have been used to fabricate high-definition plasmonic nanodevices [27, 32, 33], the chemical method for the synthesis of microscale single-crystalline aluminum plates is still missing. In this work, we employ molecular beam epitaxy (MBE) method to grow high quality single-crystalline aluminum film on top of GaAs substrate and apply focused-ion beam (FIB) milling to define plasmonic nanostructures, including periodic nanoslit arrays and nanoholes. Periodic nanoslit arrays and nanoholes are used because they are well-studied systems for characterizing surface plasmon polaritons (SPPs) and localized surface plasmon resonance (LSPR). Both linear scattering spectrum and non-linear two-photon photoluminescence (TPPL) of the nanostructures are systematically studied. The effect of high index GaAs substrate on the plasmonic resonance of aluminum nanostructures is also studied. Our work gains insight into the optical property of single-crystalline aluminum nanostructures on high-index semiconducting GaAs substrate and demonstrates the potential of single-crystalline aluminum in UV to NIR plasmonics.

**RESULTS AND DISCUSSION**

*Structural and optical property of Single-crystalline aluminum film*

Our single-crystalline aluminum film is prepared following our previously reported procedure [34]. Briefly, a GaAs buffer layer (thickness = 200 nm) is firstly grown on an undoped GaAs substrate in an ultrahigh vacuum chamber ($3\times10^{-10}$ Torr) to ensure a perfectly smooth (100) crystalline surface of GaAs. Aluminum film is then grown *in-situ* on the (100) surface of the GaAs buffer layer using a solid-source molecular beam epitaxy system (Gen II, Varian). The growing process is controlled at a constant temperature of 0 °C and the growth rate is kept at 0.05 nm/s until the thickness of the aluminum film reaches 40 nm. Figure 1(a) shows a representative scanning electron microscope (SEM) image of the surface of single-crystalline aluminum film. For comparison, SEM image of the surface of multi-crystalline aluminum film prepared by commonly used electron-gun evaporator is shown in Fig. 1(b). It is clear that the MBE grown single-crystalline aluminum film has an almost perfectly flat surface, whereas the evaporated multi-crystalline film obviously contains surface corrugation and voids. Thanks to the ultrasmooth surface, the MBE grown single-crystalline aluminum film shows higher reflectivity in the UV to NIR spectral regime compared to the multi-crystalline one (Fig. 1(c)). For spectral window below 300 nm, the reflectance of the single-crystalline film is even up to 35% higher than that of the multi-crystalline one. The dip in the reflectivity around 820 nm for both curves is due to the interband transition of aluminum.

To further compare the flatness and the film quality, we have measured the TPPL from the two aluminum films under the same excitation power (4.5 mW on sample). The principle of such examination is based on the fact that TPPL is very sensitive to and can be greatly enhanced by surface roughness and nanoscale defects because the absorption of the first photon in the TPPL process regards an intraband transition and requires strong field gradient to match the momentum [35]. Therefore, defects and surface roughness can boost the TPPL, whereas an ultrasmooth single-crystalline film, lacking hot spots and strong field gradient, cannot efficiently generate TPPL. As shown in Fig. 1(d), TPPL is indeed highly enhanced by the rough surface of the multi-crystalline aluminum film. In contrast, the TPPL from the atomically flat single-crystalline aluminum film is almost absent, confirming the perfect surface quality. From a fabrication point of view, MBE method is beneficial because it allows for precise control of film thickness. The single-crystallinity also facilitates the FIB fabrication of complex plasmonic networks containing nanoscale fine features over microscale area [27]. As can be seen in Figure 1(a), long and narrow nanogaps can be easily fabricated on the single-crystalline aluminum film without a single defect. Such high-definition structures are very difficult to fabricate by applying FIB onto multi-crystalline aluminum film because the random crystal grains and surface roughness introduce variation of the resistance to the ion-beam milling and therefore result in unpredictable structural defects.

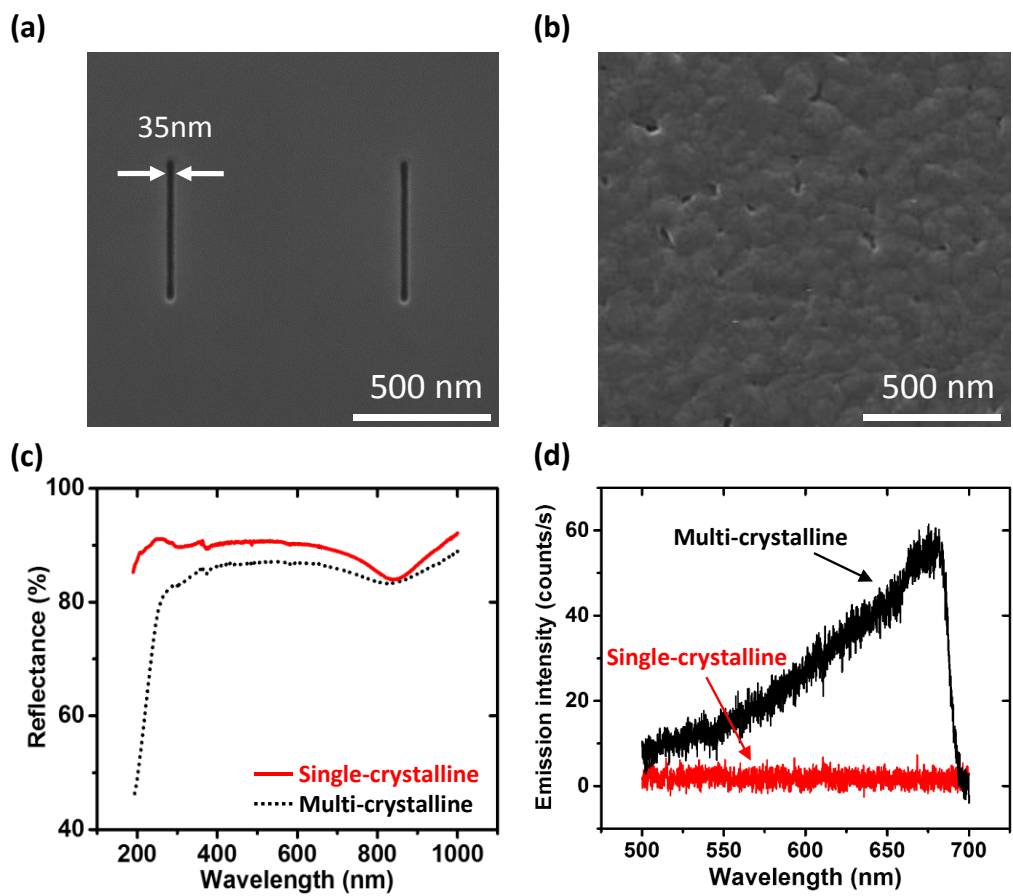

**Figure 1.** (a) Representative SEM image of the single-crystalline aluminum film prepared by MBE on GaAs substrate. The two nanogaps are created by FIB milling into the ultrasmooth aluminum film. Scale bar: 500 nm. (b) Representative SEM image of a multi-crystalline aluminum film prepared by electron-gun evaporation. The rough surface contains random grains and voids. (c) Reflectance of the single-crystalline (red solid line) and multi-crystalline (black dotted line) aluminum film. (d) TPPL spectra of single-crystalline (red) and multi-crystalline (black) aluminum film under the same illumination power (4.5 mW on sample).

*UV to NIR Surface plasmon resonance of nanoslit arrays*

To demonstrate UV to NIR surface plasmon resonance, we first examine the scattering of aluminum nanoslit arrays using a home-built dark-field microscope (Figure S1, Supporting

Information). Because our GaAs substrate is not transparent, we illuminate the sample and collect the scattered light from the same side. Figure 2(a) shows the SEM images and the dark-field scattering image of eight periodic nanoslit arrays fabricated on the single-crystalline aluminum film on top of GaAs substrate. These nanoslit arrays are marked as array #1 to array #8 with increasing periodicity from 250 nm to 600 nm in steps of 50 nm. The gap width and the length of the slits are 40 nm and 1 μm, respectively. The inter-array distance is set to 1 μm. The periodic nanoslit arrays serve as efficient couplers that provide the needed momentum for the free-space photons to excite SPPs. The coupling condition is described by Bloch equation as [36],

$$k_0 \sin\theta \pm m \cdot 2\pi/P = k_0 \cdot \sqrt{\frac{\varepsilon_m \varepsilon_d}{\varepsilon_m + \varepsilon_d}} \qquad (1)$$

, where $k_0$ is the wavenumber of light in vacuum, $\theta$ is the incident angle, $m$ is the grating order and $P$ is the periodicity. $\varepsilon_m$ and $\varepsilon_d$ are the permittivity of aluminum and the dielectric, respectively. As the periodicity of the nanoslit array decreases, the dark-field scattering images show clear evolution of color from orange to blue and eventually dark due to the low sensitivity of the color CCD. The scattering spectra of the nanoslit arrays are shown in Fig. 2(b) with the corresponding simulated spectra shown on the right panel. For the nanoslit arrays used in this work, two resonances corresponding to the fundamental (m = 1) and first higher-order (m = 2) resonance are observed within the observation window between 350 nm and 800 nm. In the simulated spectra, both resonances are clearly seen and gradually red shift as the periodicity increases, as described by equation (1). However, the fundamental resonance (m = 1) is missing in the experimental spectra. This is due to the fact that the scattering angle of the fundamental mode is the same as the incident angle. Therefore, the fundamental mode is completely blocked by the pinhole and cannot be observed (Figure S2, Supporting Information). Using a dedicated setup for signals in the UV regime, plasmonic resonance below 400 nm can be clearly observed (Figure S3, Supporting Information). In both the experimental spectra and simulated spectra, asymmetric Fano-like spectral profiles are observed. Such asymmetric line shape is due to the coupling between the broad resonance of the single slits and the relatively sharp resonance of the nanoslit array [37] (see Figure S4, Supporting Information).

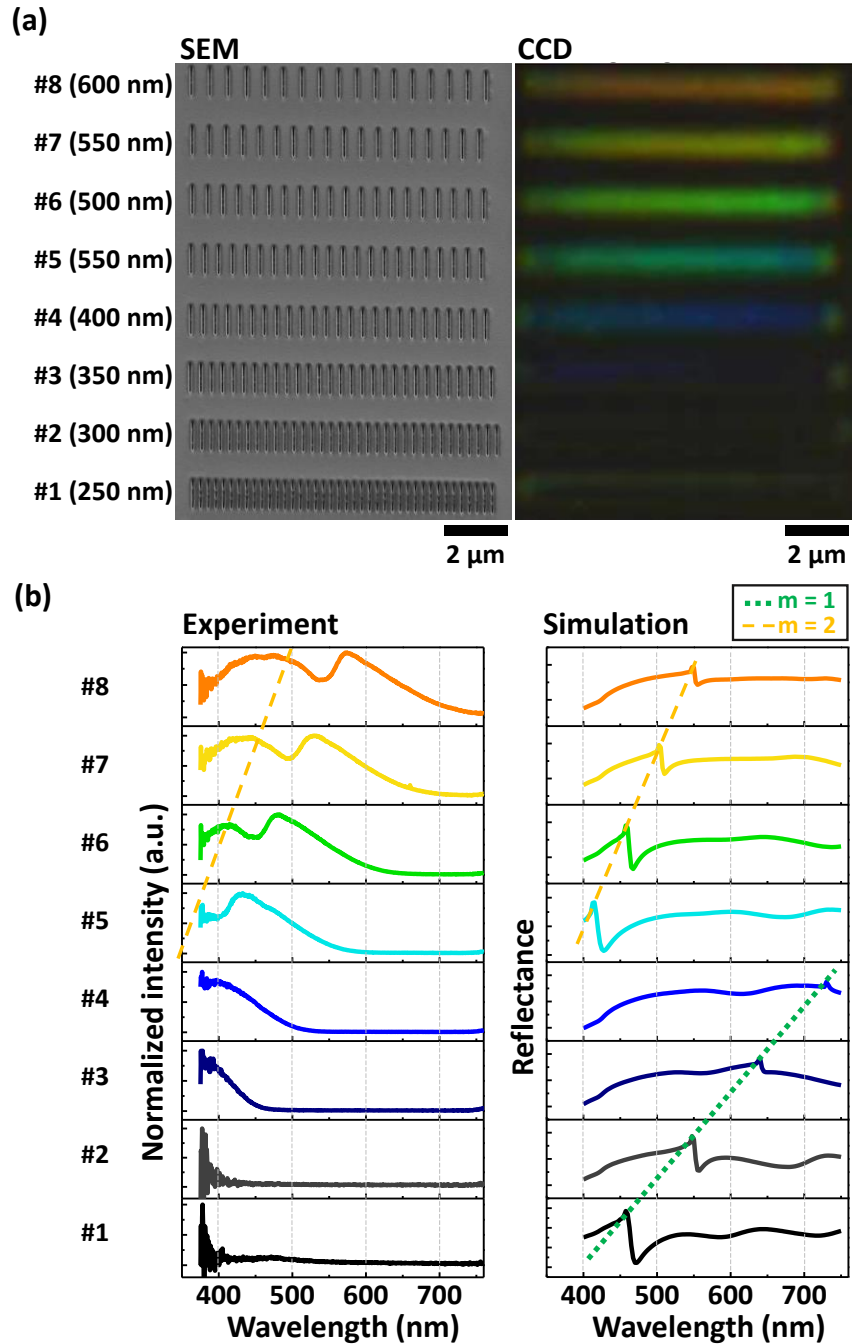

**Figure 2.** (a) Left panel: SEM image of the periodic nanoslits arrays numbered as #1 to #8 with periodicity ranging from 250 to 600 nm. Right Panel: Corresponding dark-field scattering image of the arrays. Scale bar: 2 μm. (b) The experimental dark-field scattering spectra (left) and the simulated reflectance spectra (right), showing the fundamental mode (m = 1, green dotted line) and the first higher-order mode (m = 2, orange dashed line).

*LSPR of Single Nanoholes*

Next, we examine the localized surface plasmon resonance (LSPR) of single nanoholes on our single-crystalline aluminum film. Compared to solid aluminum structures, such as nanodisks, [24, 25] nanospheres [38] and nanorods [39-41], inverse nanohole structures are easy to fabricate and is of great potential in plasmonic optical trapping [11, 42]. Figure 3(a) shows the SEM images of all the single nanoholes on the aluminum film with corresponding dark-field scattering images on the right side. For each diameter, we have fabricated a 3 × 3 array of duplicated nanoholes. The inter-hole distance is set to be 3 μm in order to reduce the cross talk between holes. The diameter of the hole is scanned from 150 to 750 nm in steps of 50 nm, resulting in thirteen different nanohole diameters, marked as #1 to #13. Their scattering images show clear red shift as the nanohole diameter increases. Dark-field scattering spectra of the nanoholes and the corresponding simulated spectra are shown in Fig. 3(b). The experimental spectra show two resonant modes in good agreement with the simulated scattering spectra. These two modes are the in-plane LSPR mode of the aluminum nanohole and the photonic mode of the vertical air hole in GaAs (Figure S5, Supporting Information). The vertical air channel is produced because FIB is milling too deep into the GaAs substrate. Therefore, the vertical "air channel" in the GaAs essentially forms an out-of-plane Fabry-Pérot (F-P) cavity for the photonic modes. Increasing the nanohole diameter shifts the in-plane LSPR mode from the UV regime (nanoholes #1 and #2) to the NIR regime (nanoholes #10 to #13), as marked with the red dotted lines in Fig 3(b). The photonic mode in the vertical GaAs nanohole (blue dashed lines in Fig. 3(b)) also shows slight red shift with increasing diameter. Although the cavity length, *i.e.* the depth of the air hole in GaAs, is rather constant for all holes, the effective index of the photonic mode slightly increases with increasing the diameter (Table S1, Supporting Information). As a result, the resonance wavelength slightly red shifts with increasing diameter. In fact, such a photonic mode exists even with the aluminum film (Figure S5, Supporting Information). The photonic mode is not seen for nanoholes with diameter smaller than 200 nm because the mode is cut off, a clear feature of photonic modes in dielectric waveguides. The photonic mode is unique for GaAs substrate and is less pronounced in low index substrate, such as glass. The presence of these two modes in one nanohole structure adds more functions to the nanohole and can be useful for applications which require interaction of two resonant modes. For example, one may trap very small particles using the enhanced optical field of the LSPR mode [11, 42] and use the photonic mode to enhance light-matter interaction.

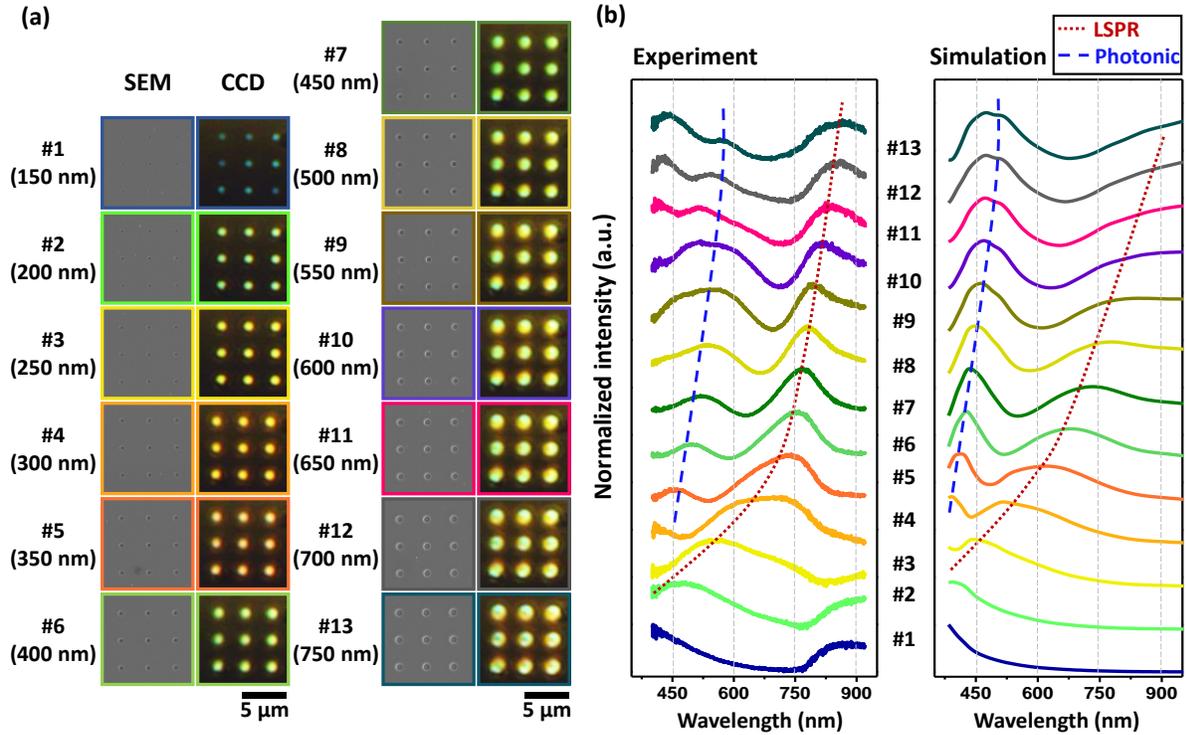

**Figure 3.** (a) The SEM and scattering images of the single nanoholes marked by #1 to #13 with diameter ranging from 150 nm to 750 nm in steps of 50 nm. Scale bars: 5 μm. (b) The experimental dark-field scattering spectra (left) and the simulated scattering spectra (right) of the nanoholes. Two modes due to the in-plane LSPR (red dotted line) mode and the Fabry-Pérot resonance of the photonic mode in the vertical air channel in the GaAs substrate (blue dashed line) are observed.

*TPPL mapping and polarization-dependent TPPL intensity*

To further understand the resonance and the photoluminescence properties of plasmonic aluminum nanoholes on GaAs substrate, we have performed TPPL mapping [27, 43, 44, 45] of single nanoholes and systematically studied the relationship between the excitation polarization and the TPPL intensity. TPPL maps are generated by plotting the TPPL intensity as a function of the position of laser focus, which is raster scanned over the nanoholes in steps of 50 nm. Because the GaAs substrate also gives strong nonlinear luminescence, we have integrated only the spectrum between 570 nm and 590 nm in order to exclude the interference from the nonlinear signals of GaAs (Figure S6, Supporting Information). The two-photon nature of the luminescence is confirmed by the quadratic dependence of the luminescence signals on the excitation power. Figure 4(a) shows the TPPL intensity as a function of the nanohole diameter. The corresponding TPPL maps for single nanoholes are displayed on top of the plot. The TPPL intensity reaches the maximum at nanohole #4 and #5 and gradually decreases as the diameter further increases, showing clear resonance-enhanced TPPL generation. As can be seen in the dark-field scattering spectra shown in Fig. 3(b), the LSPR peaks of nanohole #4, #5 and #6 best overlap with the wavelength of the excitation laser, *i.e.* 770 nm. This doubly confirms the relationship between TPPL generation and the diameter-dependent LSPR of the nanoholes. We have performed FDTD simulations to obtain the quadruplicate electric field ($|E|^4$) inside the aluminum metal in order to confirm the trend of TPPL intensity as a function of the nanohole diameter. Since TPPL is a two-photon nonlinear process, $|E|^4$ can be considered as a quantity proportional to the TPPL intensity [27, 45]. As shown in Fig. 4(a), the results obtained from the simulations (open squares linked by black solid line) agree well with the experimental results (red solid dots).

In fact, for nanoholes with diameter larger than 500 nm, they are able to support higher-order quadrupolar LSPR modes. Since our laser is tightly focused (spot size = 520 nm), the symmetry of the system is broken as we scan the laser focal spot through the nanohole and the originally dark quadrupolar LSPR mode of the large holes can be excited. The excitation of qudrupolar LSPR mode leads to the ring-shaped parrterns of nanoholes #11 to #13 in the TPPL map.

To compare the dipolar and quadrupolar resonances, we take two nanoholes with diameter of 300 nm (#4) and 750 nm (#13) as examples. They represent nanoholes with diameter smaller and larger than the size of laser focal spot. These two nanoholes support the dipolar and quadrupolar LSPR modes, respectively (Figure S7, supporting Information). The excitation power for nanohole #13 has been increased to 3 mW, *i.e.* three times larger than the power used for nanohole #4, in order to clearly visualize the TPPL pattern. Upon our laser excitation, the two nanoholes show distinct patterns in the TPPL map, as shown in Figs. 4(b) and 4(c). For the nanohole #4, a single maximum of the TPPL intensity is obtained as the laser spot is scanned through the center of the nanohole, revealing a dipolar LSPR resonance. For the nanohole #13, interestingly, a ring-shaped intensity distribution with slightly higher intensity at the upper and lower edges in y-direction is observed (Fig. 4(c)). Since our excitation is polarized in x-direction, such observation of relatively high TPPL intensity in y-direction is counter-intuitive. Typically, laser excitation with polarization perpendicular to the edge of the hole is expected to be more efficient in exciting TPPL compared to excitation with polarization parallel to the edge [46]. Therefore, one would intuitively expect that the two-lobe pattern is along the x-direction instead of y-direction. The right panels of Figs. 4(b) and 4(c) show the line-cut profiles of the experimental TPPL maps and the simulated $|E|^4$ profiles along the same x- and y-direction line. For both holes, the simulated line-cut profiles are in good agreement with the experimental ones and the counter-intuitive pattern for nanohole #13 is reproduced.

To understand this pattern, we first make sure that the counter-intuitive two-lobe pattern of the 750-nm nanohole is not due to the laser scanning direction or the structural asymmetry due to FIB milling. To this purpose, we have changed the excitation polarization to the y-direction and have observed that the two-lobe pattern changes accordingly, *i.e.* higher intensity is now found in the x-direction. This confirms that the counter-intuitive two-lobe pattern is purely determined by the polarization of the excitation. Next, we perform a series of simulations to examine the trend of $|E|^4$ as a function of the laser polarization angle with respect to the edge of the aluminum film.

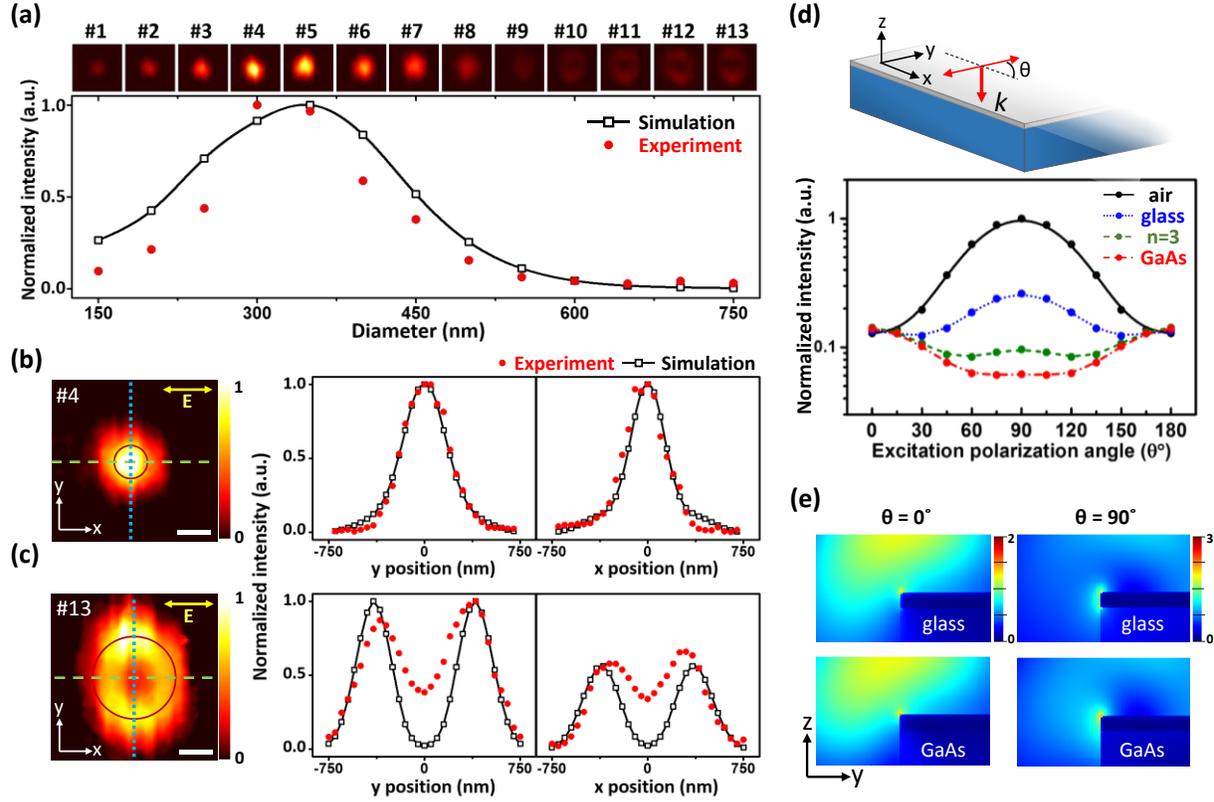

**Figure 4.** (a) Upper row: TPPL intensity maps of nanohole #1 to #13 with a scanning size of 1.5×1.5 μm$^2$. Lower row: Normalized maximum intensity in the experimental TPPL intensity maps (red dots) and the calculated $|E|^4$ (black open squares) with respect to the nanohole diameter. (b) and (c) Left panels: the intensity maps of the nanoholes with diameter 300 nm (#4) and 750 nm (#7), respectively. Scale bars are 300 nm. Right panels: Corresponding intensity profiles of the normalized experimental (red dots) and simulated intensity (black open squares) along the lines in x- (green dashed line) and y-direction (blue dotted line) cutting through the nanohole intensity maps. (d) The calculated $|E|^4$ and excitation polarization dependence on infinite straight edge with different substrates. (e) The electric near-field distributions around the edge of an aluminum film (thickness = 40 nm) on glass (top) and GaAs (bottom) under excitation with polarization along the x-direction (θ = 0°, left) and y-direction (θ = 90°, right).

Four different substrates are used in the simulations, including air, glass, a hypothetical substrate (n = 3.0) and the GaAs (n = 3.7 + 0.092i). An infinitely extended straight edge of an aluminum film (thickness = 40 nm) is used as a model structure to reflect the excitation efficiency of a polarized illumination at the film edge. The extended straight edge and the excitation geometry are depicted in the top panel of Fig. 4(d). For each substrate, we illuminate the edge of an aluminum film with a Gaussian beam (N.A. = 0.9) and simulate the $|E|^4$ in the aluminum material as a function of the excitation polarization angle ($\theta$) with respect to the x-direction ($\theta = 0°$). Figure 4(d) shows that, for different substrates, the $|E|^4$ in the aluminum film exhibits different dependence on the polarization angle. As the refractive index of the substrate increases, the $|E|^4$ inside the aluminum decreases rapidly if the excitation is polarized in the y-direction ($\theta = 90°$), *i.e.* perpendicular to the edge. The $|E|^4$ under excitation polarized along the x-axis ($\theta = 0°$), i.e. parallel to the edge, is rather insensitive to the change of substrate. For a free aluminum film in air or on top of a glass substrate, the maximum of $|E|^4$ is obtained with excitation polarization perpendicular to the film edge ($\theta = 90°$), as would be normally expected. However, for the hypothetical substrate (n = 3) and the GaAs substrate (n = 3.7 + 0.092i), the $|E|^4$ is severely suppressed when the excitation polarization is perpendicular to the edge. As a result, the relatively large $|E|^4$ and relatively pronounced TPPL are obtained when the excitation polarization is parallel to the edge. This explains the counter-intuitive two-lobe pattern observed in the TPPL pattern of the 750-nm hole. By further examining the simulated near-field distribution (Fig. 4(e)), we found that the spatial distribution of the excited near field is strongly depend on the excitation polarization. This is due to the fact that excitation with different polarization couples differently into the aluminum film. For excitation polarization perpendicular to the edge ($\theta = 90°$), the field couples more into the aluminum film compared to the case with excitation polarization parallel to the edge (θ = 0°). Therefore, the index of the substrate has relatively large influence on the field. Since GaAs substrate is highly absorbing, it damps the electromagnetic field more severely under perpendicularly polarized excitation. Consequently, the counter-intuitive two-lobe pattern is obtained. Such a property is unique for the high-index GaAs substrate and needs to be considered in the design of plasmonic structures. Here, we note that the use of single-crystalline aluminum film is the key for the observation of such counter-intuitive TPPL pattern because the ultrasmooth surface and the nearly perfect rim of the hole produce extremely low background noise (Fig. 1(d)) and therefore guarantee successful observation of the small difference in the intensity of the TPPL pattern.

Finally, we study the polarization of the TPPL from aluminum. Aluminum has been reported to preserve the excitation polarization [39]. In this work, we perform polarization analysis of TPPL using the nanohole #4 (diameter = 300 nm). Since this nanohole is radially symmetric, we can rule out any possible contribution form the anisotropy of structural geometry to the fluctuation of TPPL intensity. Figure 5(a) shows the emission polar plot of the TPPL from nanohole excited with illumination polarized along the x-direction (θ = 0°). A degree of linear polarization (DoLP) of about 0.38 is obtained for the emission, suggesting that the TPPL from the aluminum hole is preferably polarized along the excitation polarization. By plotting the emission polarization-dependent intensity as a function of the excitation polarization, we have obtained a polarization correlation plot, as shown in Fig. 5(b). It can be seen that the emission polarization of TPPL well follows the polarization of the excitation, giving direct evidence that the aluminum TPPL preserves the polarization of the excitation. It is worth noting that the total intensity of the TPPL does not vary with the excitation polarization, suggesting that the TPPL yield of aluminum is independent of the crystallinity of our single-crystalline aluminum film.

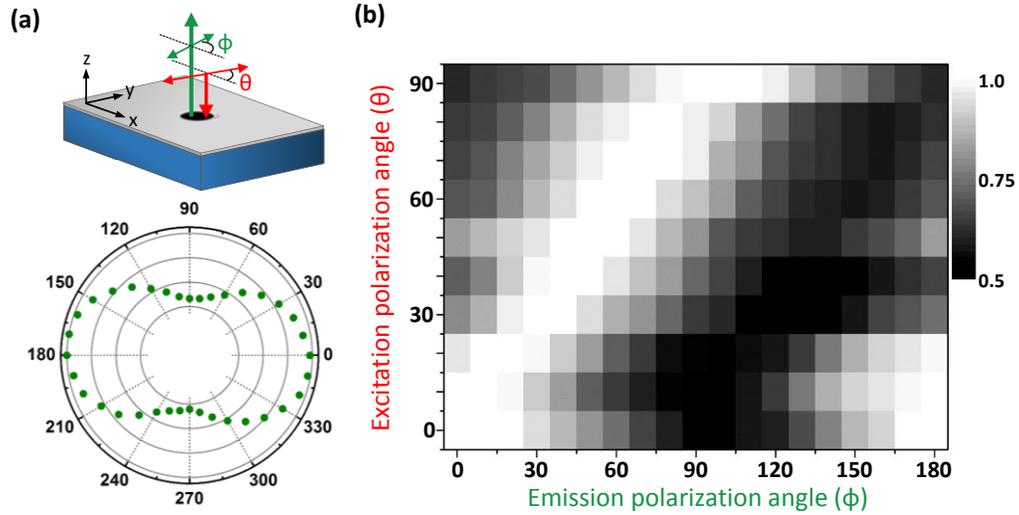

**Figure 5.** (a) Top: A schematic diagram of the excitation (θ) and emission polarization (φ). Bottom: A polar plot showing the emission polarization of the TPPL from the nanohole #4 (diameter = 300 nm) under x-polarized excitation (θ = 0°). The DoLP of the TPPL is about 0.38. (b) Polarization correlation plot of the TPPL from aluminum nanohole #4.

**Conclusion**

We have presented single-crystalline aluminum nanostructures on aluminum film grown with MBE method on top of a high-index GaAs substrate. We demonstrate clear plasmonic resonance in the UV to NIR spectral regime. With TPPL intensity mapping, we found counter-intuitive TPPL pattern that shows relatively high TPPL yield with excitation polarization parallel to the edge of the aluminum film. The counter-intuitive pattern is found to relate to the high index of the GaAs substrate. With our single-crystalline aluminum film and the radially symmetric nanohole structure, we confirm that the polarization of TPPL from aluminum depends mainly on the excitation polarization but not on the crystallinity of the metal or substrate. Aluminum nanohole structures can be easily fabricated and are useful for optical trapping or manipulation. High quality single-crystalline metallic film is of great potential for the realization of complex integrated plasmonic optical nanocircuits [5, 27] and plasmonic nanolasers [31]. We anticipate broadband and CMOS compatible plasmonic nanostructures and nanophotonic devices operating in the UV to NIR regime using single-crystalline aluminum film on semiconductor.

MATERIALS AND METHODS

*Dark-field microscope*

A home-built confocal dark-field microscope is used to obtain the scattering spectra of the nanostructures (Figure S1, Supporting Information). Briefly, a ring-shape non-polarized broadband white light (HAL 100 illuminator with quartz collector, Zeiss) is collimated and concentrated onto the sample plane by a microscope objective. The scattered light is then collected by the same objective and aligned by an achromatic lens onto the imaging CCD or the entrance slit of a UV-VIS spectrometer (SR-303i-A with DU401A-BV CCD, Andor) for spectral analysis. The straight scattering of the excitation is blocked by a beam stop with a small pinhole (diameter = 4 mm) at the center. Two sets of optics optimal for the signals in the UV and visible windows are used in order to obtain optimal detection efficiency. For the clarity of presentation, results obtained with the visible optics set are presented in the main text and the results from the UV set are in the Fig. S3 in the Supporting Information. Details of the two sets of optics are summarized in Table 1. All spectra have been normalized to the wavelength-dependent quantum efficiency of the CCD detector.

**Table 1.** Details of the optics used for the detection of UV and visible signals.

|  | wavelength range (nm) | microscope objective | achromatic lens | reflection mirrors | spectrometer grating settings |
|---|---|---|---|---|---|
| UV | 300 – 600 | LMU-40X-UVB air, N.A.=0.5, Thorlabs | NUDL-50-300P, SIGMA KOKI | Aluminum mirrors PF10-03-F01, Thorlabs | 300 lines/mm Blaze at 300 nm |
| VISIBLE | 450 – 850 | MPlanApo 60X air, N.A. = 0.9, Olympus | AC254-300-A-ML f = 300 mm, Thorlabs | Silver mirrors PF10-03-P01, Thorlabs | 150 lines/mm Blaze at 800 nm |

*Setup for TPPL mapping and spectral analysis*

A home-built laser scanning microscope is used for TPPL mapping. The setup is depicted in Fig. S8 of Supporting Information. To excite the TPPL, laser pulses from a Ti:sapphire laser (center wavelength = 770 nm, pulse duration = 140 fs, repetition = 80 MHz, Chameleon Ultra II) are filtered by a long pass filter (FEL0750, Thorlabs), reflected by a dichroic mirror (FF720-SDi01, Semrock), and focused onto the sample plane by a near-infrared microscope objective (100X NeoSPlan NIC N.A. = 0.9, Olympus). A half-wave plate (AHWP10M-980, Thorlabs) is used to control the excitation polarization prior to the objective. The generated TPPL is collected by the same objective, passing through the dichroic mirror, filtered by a short pass filter ($\lambda$ < 690 nm, XVS0690, Asahi) and aligned into the entrance slit of the spectrometer (Acton SP2750, Princeton Instrument). To avoid intensity difference due to the discriminative grating efficiency on the polarization, we have used a second half-wave plate (AHWP10M-600, Thorlabs) and a linear polarizer to control the polarization of the emission prior to the spectrometer so that the spectral analysis of TPPL is performed under identical grating efficiency. The averaged excitation power is kept at 1 mW and the exposure time of the CCD equipped on the spectrometer is 1.0 second. The sample is mounted on a piezo stage with a close-loop feedback control (Physik Instrumente P-611.3S stage and E-664 controller). Spatial-resolved TPPL mapping images are obtained by scanning the sample position (step size: 50 nm) and plot the integrated TPPL intensity (integration spectral window: 570-590 nm) with respect to the excitation position.

*Numerical Simulations*

Numerical simulations are performed with finite-difference time-domain method (FDTD Solutions, Lumerical Solutions). To obtain the far-field scattering spectra of the periodic nanoslits, we have simulated the reflectance of the nanoslit array (Figure S9, Supporting Information). The incident angle of the plane wave source is set to be 25 and 55 degree to mimic the experimental incident conditions determined by the numerical aperture of the objectives for UV and visible light, respectively. Bloch boundary condition is used for periodic structures. For single-nanoholes, the total-field scattered-field (TFSF) source with wavelength ranging from 300 nm to 900 nm is used to obtain the scattering spectra. The source is set to be normally incident because the LSPRs of single nanoholes are independent of the incident angle. Perfectly matched layer (PML) boundaries are placed at least 1 µm away from the nanoholes. The scattering spectra are obtained by integrating the Poynting vector over the area of a two-dimensional monitor, which is 600 nm above the nanostructure. All simulated spectra have been normalized to the source spectrum. To simulate the generation efficiency of the TPPL as a function of excitation position, we have used a tightly focused Gaussian source synthesized by a thin lens setting (N.A. = 0.9 approximated by 200 plane waves). Square of the simulated electric field intensity $|E|^4$ is then integrated over the whole volume of metal to obtain a quantity proportional to the TPPL intensity. For all structures, the aluminum film with thickness of 40 nm, including an oxide layer of 2 nm and an aluminum layer of 38 nm, is placed on the GaAs substrate. Rounded corners (radius of curvature = 5 nm) are used to mimic the actual geometry of the nanostructures fabricated by FIB milling. The dielectric functions of Al, $Al_2O_3$ and GaAs are modeled with experimental data from Palik [47].

## ASSOCIATED CONTENT

**Supporting Information**. Dark-field scattering setup, Objective collection efficiency, UV spectra, Spectral profile analysis, TPPL spectrum, Mode analysis, TPPL mapping setup, FDTD settings. This material is available free of charge via the Internet at http://pubs.acs.org.

## AUTHOR INFORMATION

**Corresponding Author**

*Email: jshuang@mx.nthu.edu.tw

**Note**

The authors declare no competing financial interest.

## ACKNOWLEDGEMENT

Support from the Ministry of Science and Technology of Taiwan under Grant Nos. NSC-101-2113-M-007-002-MY2, NSC-101-2628-E-009-MY3, and MOST-103-2113-M-007-004-MY3 are gratefully acknowledged. S.D.L. acknowledges the financial support from MOST and from ATU program of MOE in Taiwan. The equipment support from CNST and NFC at NCTU is appreciated. J.S.H. thanks the support from the Center for Nanotechnology, Materials Sciences, and Microsystems at National Tsing Hua University.

## REFERENCES

[1]. Biagioni, P.; Huang, J.-S.; Hecht, B. Nanoantennas for Visible and Infrared Radiation. *Rep. Prog. Phys.* **2012**, *75*, 024402.
[2]. Novotny, L.; van Hulst, N. Antennas for Light. *Nat. Photon.* **2011**, *5*, 83–90.
[3]. Ebbesen, T. W.; Genet, C.; Bozhevolnyi, S. I. Surface-Plasmon Circuitry. *Phys. Today* **2008**, *61*, 44−50.
[4]. Hung, Y.-T.; Huang, C.-B.; Huang, J.-S. Plasmonic Mode Converter for Controlling Optical Impedance and Nanoscale Light-matter Interaction. *Opt. Express* **2012**, *20*, 20342–20355.

[5]. Dai, W.-H.; Lin, F.-C.; Huang, C.-B.; Huang, J.-S. Mode conversion in high-definition plasmonic optical nanocircuits. *Nano Lett.* **2014**, *14*, 3881–3886.
[6]. Liu, W.-L.; Lin, F.-C.; Yang, Y.-C.; Gwo, Shangjr; Huang, M. H.; Huang, J.-S. The influence of shell thickness of Au@TiO2 core-shell nanoparticles on plasmonic enhancement effect in dye-sensitized solar cells. *Nanoscale* **2013**, 5, 7953–796.
[7]. Stewart, M. E.; Anderton, C. R.; Thompson, L. B.; Maria, J.; Gray, S. K.; Rogers, J. A.; Nuzzo, R. G. Nanostructured Plasmonic Sensors. *Chem. Rev.* **2008**, *108*, 494–521.
[8]. N.; Hall, W. P.; Lyandres, O.; Shah, N. C.; Zhao, J.; Duyne, R. P. V. Biosensing with plasmonic nanosensors. *Nat. Mater.* **2008**, *7*, 442– 453.
[9]. Lin, Daniel; Huang, J.-S. Slant-gap plasmonic nanoantennas for optical chirality engineering and circular dichroism enhancement. *Opt. Express* **2014**, *22*, 7434–7445.
[10]. Pang, Y.; Gordon, R. Optical Trapping of a Single Protein. *Nano Lett.* **2012**, *12*, 402−406
[11]. Chen, K.-Y.; Lee, A.-T.; Hung, C.-C.; Huang, J.-S.; Yang, Y.-T. Transport and trapping in two-dimensional nanoscale plasmonic optical lattice. *Nano Lett.* **2013**, *13,* 4118–4122.
[12]. Hogan, N. J.; Urban, A. S.; Ayala-Orozco, C.; Pimpinelli, A., Nordlander, P.; Halas, N. J. Nanoparticles Heat through Light Localization. *Nano Lett.* **2014**, *14*, 4640–4645.
[13]. Sobhani, A.; Knight, M. W.; Wang, Y.; Zheng, B.; King, N. S.; Brown, L. V.; Fang, Z.; Nordlander, P.; Halas, N. J. Narrowband photodetection in the near-infrared with a plasmon-induced hot electron device. *Nat. Commun.* **2013**, *4*, 1643.
[14]. Shrestha,V. R.; Lee, S-S. ; Eun-Soo Kim, E-S.; Choi, D-Y. Aluminum Plasmonics Based Highly Transmissive Polarization-Independent Subtractive Color Filters Exploiting a Nanopatch Array. *Nano Lett.* **2014**, Article ASAP.
[15] Sun, Y.; Xia, Y. Shape-Controlled Synthesis of Gold and Silver Nanoparticles. *Science* **2002**, *298*, 2176–2179.
[16] Chiu, C.-Y.; Chung, P.-J.; Lao, K.-U.; Liao, C.-W.; Huang, M. H. J. Facet-Dependent Catalytic Activity of Gold Nanocubes, Octahedra, and Rhombic Dodecahedra toward 4-Nitroaniline Reduction. *Phys. Chem. C* **2012**, *116*, 23757−23763.
[17]. Vakil, A.; Engheta, N. Transformation optics using graphene. *Science* **2011**, *332*, 1291–1294.
[18]. Fei, Z. et al. Electronic and plasmonic phenomena at graphene grain boundaries. *Nat. Nanotechnol.* **2013,** *8*, 821–825.
[19]. Boltasseva, A.; Atwater, H. A. Low-Loss Plasmonic Metamaterials. *Science* **2011**, *331*, 290.
[20]. Naik, G. V.; Schroeder, J. L.; Ni, X.; Kildishev, A. V.; Sands, T.D.; Boltasseva, A. Titanium nitride as a plasmonic material for visible and near-infrared wavelengths. *Opt. Mater. Express* **2012**, *2 (4)*, 478−489.
[21] Kinsey, N.; Ferrera, M.; Naik, G. V.; Babicheva, V. E.; Shalaev, V. M.; Boltasseva, A.; Experimental demonstration of titanium nitride plasmonic interconnects. *Opt. Express* **2014**, *22(10)*, 12238–12247.
[22]. Ray, K.; Chowdhury, M. H.; Lakowicz, J. R. Aluminum nanostructured films as substrates for enhanced fluorescence in the ultraviolet-blue spectral region. *Anal. Chem.* **2007**, *79*, 6480–6487.
[23]. Mahdavi, F.; Blair, S. Nanoaperture Fluorescence Enhancement in the Ultraviolet. *Plasmonics* **2010**, *5*, 169−174.
[24]. Langhammer, C.; Schwind, M.; Kasemo, B.; Zoric, I. Localized Surface Plasmon Resonances in Aluminum Nanodisks. *Nano Lett.* **2008**, *8*, 1461−1471.

[25]. Knight, M. W.; King, N. S.; Liu, L.; Everitt, H. O.; Nordlander, P.; Halas, N. J. Aluminum for plasmonics. *ACS Nano* **2014**, *8*, 834−840.
[26]. Kuttge, M.; Vesseur, E. J. R.; Verhoeven, J.; Lezec, H. J.; Atwater, H. A.; Polman, A. Loss Mechanisms of Surface Plasmon Polaritons on Gold Probed by Cathodoluminescence Imaging Spectroscopy. *Appl. Phys. Lett.* **2008**, *93*, 113110.
[27]. Huang, J.-S.; Callegari, V.; Geisler, P.; Brüning, C.; Kern, J.; Prangsma, J. C.; Wu, X.; Feichtner, T.; Ziegler, J.; Weinmann, P.; Kamp, M.; Forchel, A.; Biagioni, P.; Sennhauser, U.; Hecht, B. Atomically Flat Single-crystalline Gold Nanostructures For Plasmonic Nanocircuitry. *Nature Comm.* **2010**, *1*, 150.
[28] Lu, Y. J.; Kim, J.; Chen, H. Y.; Wu, C. H.; Dabidian, N.; Sanders,C. E.; Wang, C. Y.; Lu, M. Y.; Li, B. H.; Qiu, X. G.; Chang, W. H.; Chen, L. J.; Shvets, G.; Shih, C. K.; Gwo, S. Plasmonic Nanolaser Using Epitaxially Grown Silver Film. *Science* **2012**, *337*, 450-453.
[29]. Levine, I.; Yoffe, A.; Salomon, A.; Li, W.; Feldman, Y.; Cahen, D.; Vilan, A. Epitaxial Two Dimensional Aluminum Films on Silicon (111) by Ultra-Fast Thermal Deposition. *J. Appl. Phys.* **2012**, *111*, 124320.
[30]. Stefaniuk, T.; Wróbel, P.; Ciesielski, A.; Szoplik, T. Fabrication of Smooth Al Nanolayers at Different Temperatures. *ICTON*. **2013**, *Tu.P.26*.
[31] Aspnes, D. E.; Kinsbron, E.; Bacon, D. D. Optical Properties of Au: Sample Effects. *Phys. Rev. B* **1980**, *21*, 3290−3299.
[32]. Chen, W.-L.; Lin, F.-C.; Lee, Y.-Y.; Li, F.-C.; Chang, Y.-M.; Huang, J.-S. The Modulation Effect of Transverse, Antibonding, and Higher-Order Longitudinal Modes on the Two-Photon Photoluminescence of Gold Plasmonic Nanoantennas. *ACS Nano* **2014**, *8*, 9053-9062.
[33]. Chang, C.-W.; Lin, F.-C.; Chiu, C.-Y.; Su, C.-Y.; Huang, J.-S.; Perng, T.-P.; Yen, T.-J., HNO3‑Assisted Polyol Synthesis of Ultralarge Single-Crystalline Ag Microplates and Their Far Propagation Length of Surface Plasmon Polariton. *ACS Appl. Mater. Interfaces* **2014**, *6*, 11791-11798.
[34]. Lin, S.-W.; Wu, J.-Y.; Lin, S.-D.; Lo, M.-C.; Lin, M.-H.; Liang, C.-T., Characterization of Single-Crystalline Aluminum Thin Film on (100) GaAs Substrate. *Jpn. J. Appl. Phys.* **2013**, *52*, 045801.
[35] Beversluis, M. R.; Bouhelier, A.; Novotny, L. Continuum Generation from Single Gold Nanostructures Through Near-field Mediated Intraband Transitions. *Phys. Rev. B* **2003**, *68*, 115433.
[36]. Barnes, W. L.; Murray, W. A.; Dintinger, J.; Devaux, E.; Ebbesen, T. W. Surface Plasmon Polaritons and Their Role in the Enhanced Transmission of Light Through Periodic Arrays of Subwavelength Holes in a Metal Film. *Phys. Rev. Lett.* **2004**, *92*, 107401.
[37]. Lee, K. L.; Chen, P. W.; Wu, S. H.; Huang, J. B.; Yang, S. Y.; Wei, P. K. Enhancing Surface Plasmon Detection Using Template-Stripped Gold Nanoslit Arrays on Plastic Films. *ACS Nano* **2012**, *6*,2931–2939.
[38]. Sanz, J. M.; Ortiz, D.; Alcaraz de la Osa, R.; Saiz, J. M.; González, F.; Brown, A. S.; Losurdo, M.; Everitt, H. O.; Moreno, F.UV Plasmonic Behavior of Various Metal Nanoparticles in the Near- and Far-Field Regimes: Geometry and Substrate Effects. *J. Phys. Chem. C* **2013**,*117*, 19606–19615.
[39]. Castro-Lopez, M.; Brinks, D.; Sapienza, R.; van Hulst, N. F. Aluminum for Nonlinear Plasmonics: Resonance-Driven Polarized Luminescence of Al, Ag, and Au Nanoantennas. *Nano Lett.* **2011**, *11*, 4674−4678.

[40]. Knight, M. W.; Liu, L. F.; Wang, Y. M.; Brown, L.; Mukherjee, S.; King, N. S.; Everitt, H. O.; Nordlander, P.; Halas, N. J. Aluminum plasmonic nanoantennas. *Nano Lett.* **2012**, *12*, 6000−6004.

[41]. Schwab, P. M.; Moosmann, C.; Wissert, M. D.; Schmidt, E. W.-G.; Ilin, K. S.; Siegel, M.; Lemmer, U.; Eisler, H.-J. Linear and Nonlinear Optical Characterization of Aluminum Nanoantennas. *Nano Lett.* **2013**, *13*, 1535–1540.

[42]. Berthelot, J.; Acˊimovicˊ, S. S.; Juan, M. L.; Kreuzer, M. P.; Renger, J.; Quidant, R. Three-dimensional manipulation with scanning near-field optical nanotweezers. *Nat. Nanotechnol.* **2014**, *9*, 295–299.

[43]. Imura, K.; Nagahara, T.; Okamoto, H. Near-Field Two-Photon-Induced Photoluminescence from Single Gold Nanorods and Imaging of Plasmon Modes. *J. Phys. Chem. B* **2005**, *109*, 13214−13220.

[44]. Viarbitskaya, S.; Teulle, A.; Marty, R.; Sharma, J.; Girard, C.; Arbouet, A.; Dujardin, E. Tailoring and Imaging the Plasmonic Local Density of States in Crystalline Nanoprisms. *Nat. Mater.* **2013**, *12*,426−432.

[45]. Huang, J.-S.; Kern, J.; Geisler, P.; Weinmann, P.; Kamp, M.; Forchel, A.; Biagioni, P.; Hecht, B. Mode Imaging and Selection in Strongly Coupled Nanoantennas. *Nano Lett.* **2010**, *10*, 2105-2110.

[46]. Xu, T.; Wu, Y.-K.; Luo, X.; Guo, L. J. Plasmonic Nanoresonators for High-Resolution Colour Filtering and Spectral Imaging. *Nat. Commun.* **2010**, *1*, 59.

[47]. Palik, E. D. Handbook of Optical Constants of Solids; Academic Press: Orlando, FL, **1985**; Vol. I.

# Supporting Information

# Single-crystalline Aluminum Nanostructures on Semiconducting GaAs Substrate for Ultraviolet to Near-infrared Plasmonics

Hsuan-Wei Liu,[1] Fan-Cheng Lin,[2] Shi-Wei Lin,[1] Jau-Yang Wu,[1]

Sheng-Di Lin[1] and Jer-Shing Huang [2,3,*]

*[1] Department of Electronics Engineering, National Chiao Tung University, Hsinchu 30010, Taiwan*

*2 Department of Chemistry, National Tsing Hua University, Hsinchu 30013, Taiwan*

*3. Frontier Research Center on Fundamental and Applied Sciences of Matters, National Tsing Hua University, Hsinchu 30013, Taiwan*

AUTHOR ADDRESS:  jshuang@mx.nthu.edu.tw

*Dark-field scattering setup*

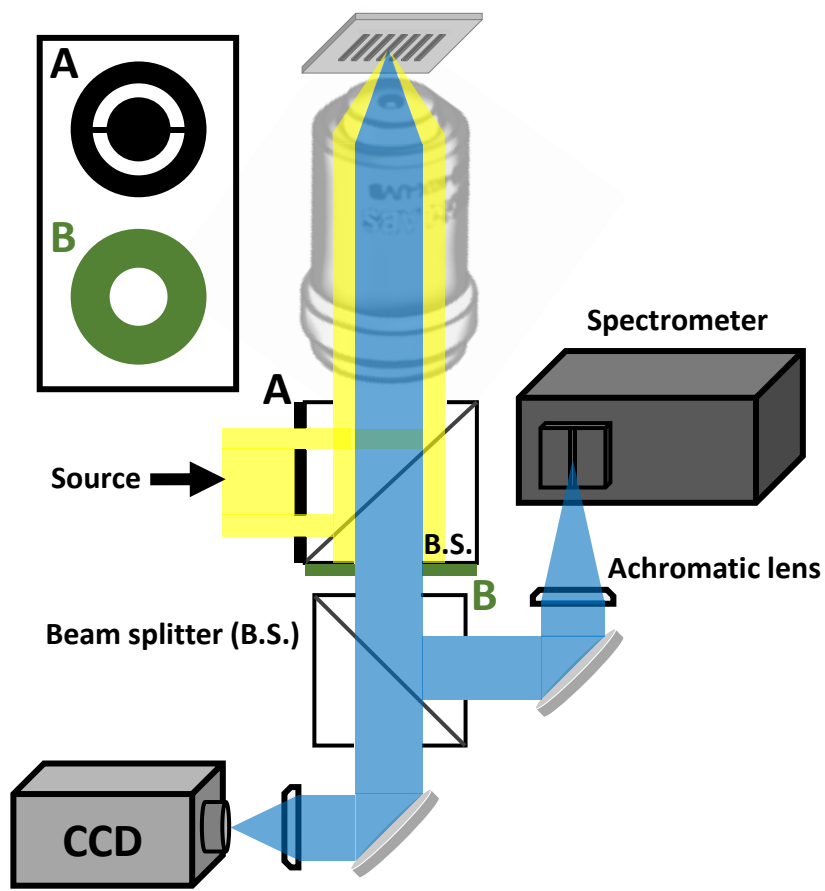

**Figure S1.** Dark-field microscope setup for scattering images and spectra. To obtain the dark-field image, the cross section of the white light source (yellow beam) is shaped into a ring by a ring-shaped aperture (black) located in front of the beam-splitter. The reflected excitation source is stopped by a beam stop (green) with a hole at the center, which allows only the scattered signal (blue beam) to pass and reach the CCD camera and spectrometer for imaging and spectral analysis.

*Objective collection efficiency*

Figure S2(a) shows the schematic diagram of the excitation and collection beam path in the objective of our dark-field microscope. The beam path of the incident and reflected white light illumination is marked by a yellow band and the scattered light from the plasmonic resonances, m = 1 and m = 2, are marked in green and orange solid lines, respectively. For each resonance order and periodicity, we can use equation (1) in the main text to analytically calculate the incident angle for exciting SPPs as a function of vacuum wavelength. The results are shown in Fig. S2(b). The incident angle of the white light illumination is marked with the yellow band with a small spread and the resonance order m = 1, m = 2 and m = 3 are marked with the solid lines in green, orange and brown color, respectively. The intersections of the yellow band (illumination) and the solid lines (resonance) thus point to the wavelengths of the allowed resonance. Because equation (1) describes not only the in-coupling condition but also the out-coupling one, the relationship between the out-coupling angle and the resonance wavelength also follows the curves in Fig S2(b). Therefore, for a specific excitation wavelength, the SPP will couple back to the free space in discrete angles depending on the resonance order. If the same resonance order is used both for in- and out-coupling, the out-coupling angle would be the same as the incident angle. In this case, the light will not pass the pinhole but be blocked by the beam stop. However, if different order is used for in- and out-coupling, the scattered light might have the chance to pass through the hole and be detected. For example, if the light couples to the nanoslit array with the m = 2 mode and the excited SPPs couple back to the far field with the m = -1 mode, the light will be scattered into the free space at an angle smaller than the incident angle. In this case, the scattered light can go through the hole and reach the CCD and spectrometer, as depicted in Fig. S2(a). An example of such case is the nanoslits of period = 500 nm, as shown in the 2$^{nd}$ plot from the left in the lower row in Fig. S2(b). The array assists the in-coupling of 480 nm light via the m = 2 resonance, *i.e.* the yellow band intercepts the m = 2 line at 480 nm (marked with the orange solid line with an arrow). The excited SPPs then couple back to far-field via the m = -1 resonance at an angle of about 7 degree, marked with the orange dotted line. In the same plot, the yellow band intercepts the m = 1 resonance at 920 nm (marked with the green solid line with an arrow), meaning the excitation of SPPs. However, with this 500-nm periodicity, there is no other intercepting point. Therefore, the 920-nm SPPs can only be scattered back to the far field at the angle equal to the incident one. Out-coupled light in this angle is blocked by the beam stop and the m = 1 resonance is thus not observed in the experimental spectra.

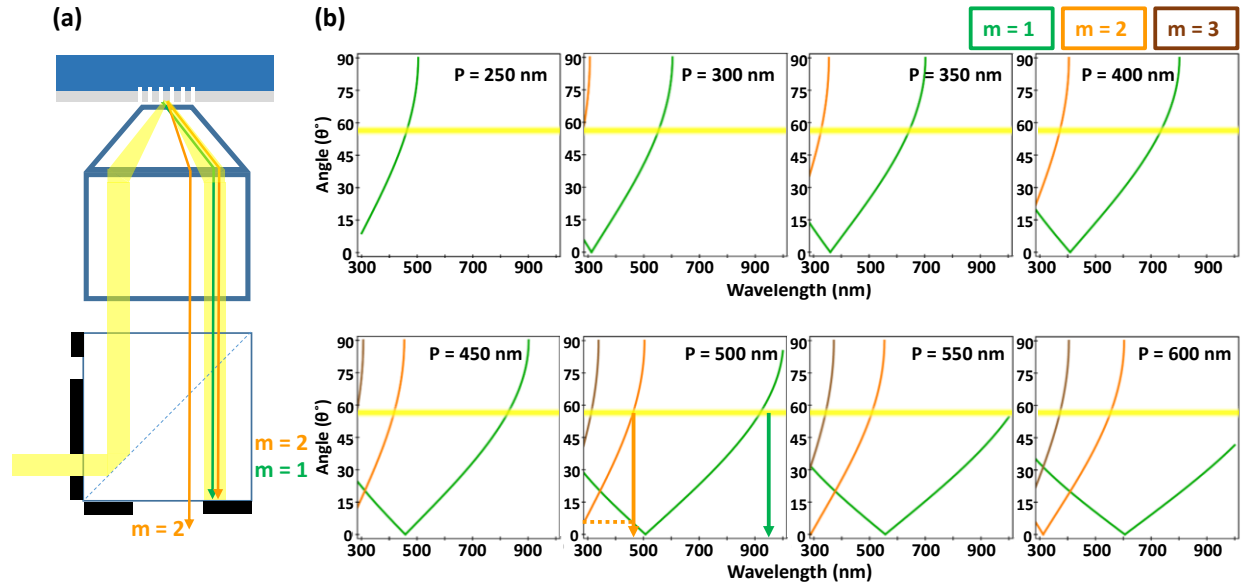

**Figure S2.** (a) The schematic of the incident (yellow band) and the scattered beam paths in the objective of the dark-field microscope setup. The green and orange solid lines are the slowed scattering beam path for the m =1 and m = 2 resonance, respectively. Only half of the excitation beam is depicted for clarity. (b) Calculated results of the in- and out-coupling angle as a function of vacuum wavelength. The incident angle of the white light illumination is marked with the yellow band and the resonance order m = 1, m = 2 and m = 3 are marked with the solid lines in green, orange and brown color, respectively. The intersections of the yellow band (illumination) and the solid lines (resonance) thus point to the wavelengths of the allowed resonance.

*UV spectra*

We have used a dedicated optics set to obtain the scattering spectra of the nanostructures in the UV regime. Details of the optics are described in Table 1 in the main text. The scattering spectra of the nanoslits obtained with the UV optics set are shown in Fig. S3(a) along with the corresponding simulated spectra. The scattering images and spectra of the nanoholes are shown in Fig. S3(b). As the periodicity of the nanoslits decreases, the higher order resonance ($m = 2$) clearly shifts from the near visible ($\lambda = 400$ nm) to the UV regime. Such resonance in the UV range is a unique character of aluminum plasmonics. Compared to the spectra in Fig. 2(b) in the main text, a systematic blue shift of all peaks is observed in Fig. S3(a). This is because the numerical aperture of the objective used in the UV optics set (N.A. = 0.5) is significantly smaller than that of the objective used in the visible optics set (N.A. = 0.9). Small numerical aperture means small incident angle of the ring-shaped illumination. According to equation (1), for a fixed material, periodicity and resonance order, decreasing incident angle would reduce the resonant wavelength and thus lead to the systematic blue shift. The difference of the incident angle due to different numerical aperture of the objective has been taken into account in the simulations and the systematic blue shift is reproduced in the simulated spectra.

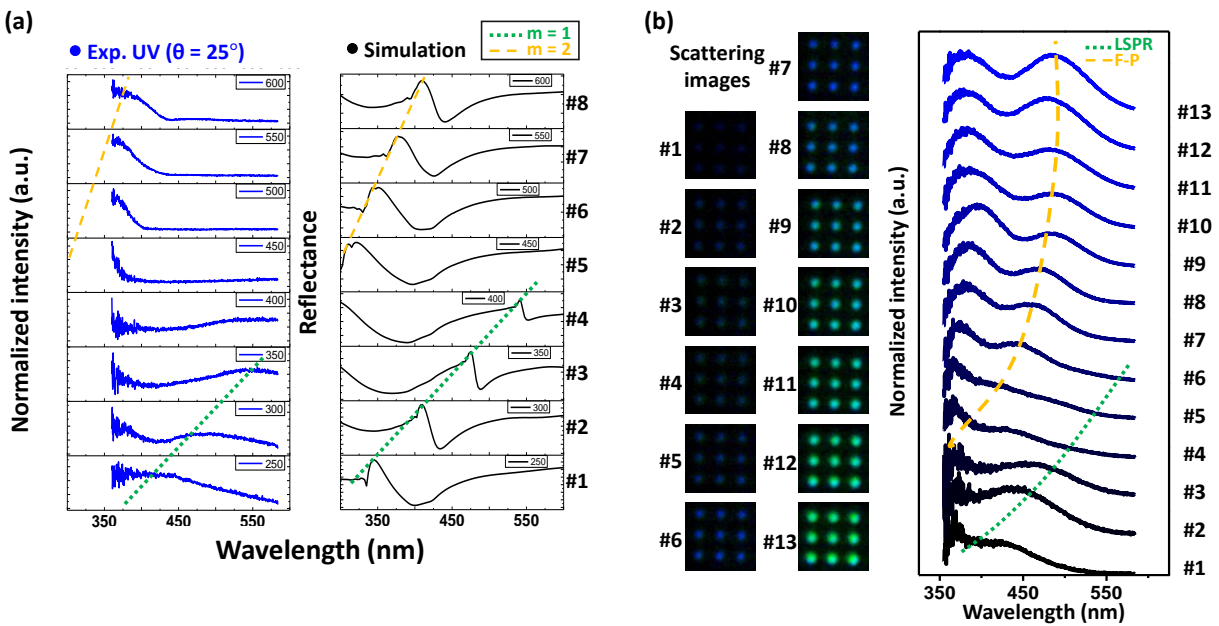

**Figure S3.** (a) Scattering (left) and simulated (right) spectra of the periodic nanoslits. (b) The scattering images (left) and spectra (right) of single nanoholes in the UV regime. All experimental spectra are obtained with the dedicated optics set for UV spectral range.

*Spectral profile analysis*

**(a) Fano-like resonance of nanoslit arrays**

In Fig. 2(b) in the main text, the nanoslit arrays with periodicity ranging from 450 to 600 nm show pronounced asymmetric Fano resonance spectral profiles for the m = 2 resonance. The Fano line shape stems from the coupling of the broad resonance of a single slit and the sharp in-plane SPP resonance of the slit array. Here, we take the nanoslit array with P = 500 nm as an example to discuss the origin of the Fano line shape. The simulated reflectance spectrum of the nanoalit array is showed in Fig. S4(a). The spectrum of one single slit is shown in Fig. S4(b). Figures S4(c) and S4(d) show the near-field distributions of $|E_x|$, $|E_y|$ and $|E_z|$ at the Fano peak ($\lambda$ = 457.5 nm, marked with gree arrow in Fig. S4(a)) and the Fano dip ($\lambda$ = 467.5 nm, marked with red arrow in Fig. S4(a)). As can be seen in the near-field distributions, the m = 2 mode corresponds to the in-plane resonance on the air/Au interface. The dip, on the other hand, stems from the coupling of the surface resonance mode with the out-of-plane cavity resonance in the vertical gap. Therefore, the field gets into the gap and the spectrum shows a dip at this frequency. For a single nanoslit on the aluminum film, the in plane m = 2 mode does not exist and the only spectral feature is the dip due to the out-of-plane cavity resonance. Therefore, there is only one broad dip around 400 nm, as shown in Fig. S4(b). Increasing the periodicity of the nanoslits shifts only the in-plane plasmonic resonance but not the out-of-plane cavity resonance, which is determined by the thickness of the film. For this reason, the asymmetric Fano line shape seen in the simulated spectra in Fig. 2(b) in the main text becomes pronounced as the in-plane resonance is tuned by the periodicity to 400 nm. Such evolution of the Fano line shape is also observed for the m = 1 resonance in the simulated spectra (right panel of Fig. 2(b) in the main text).

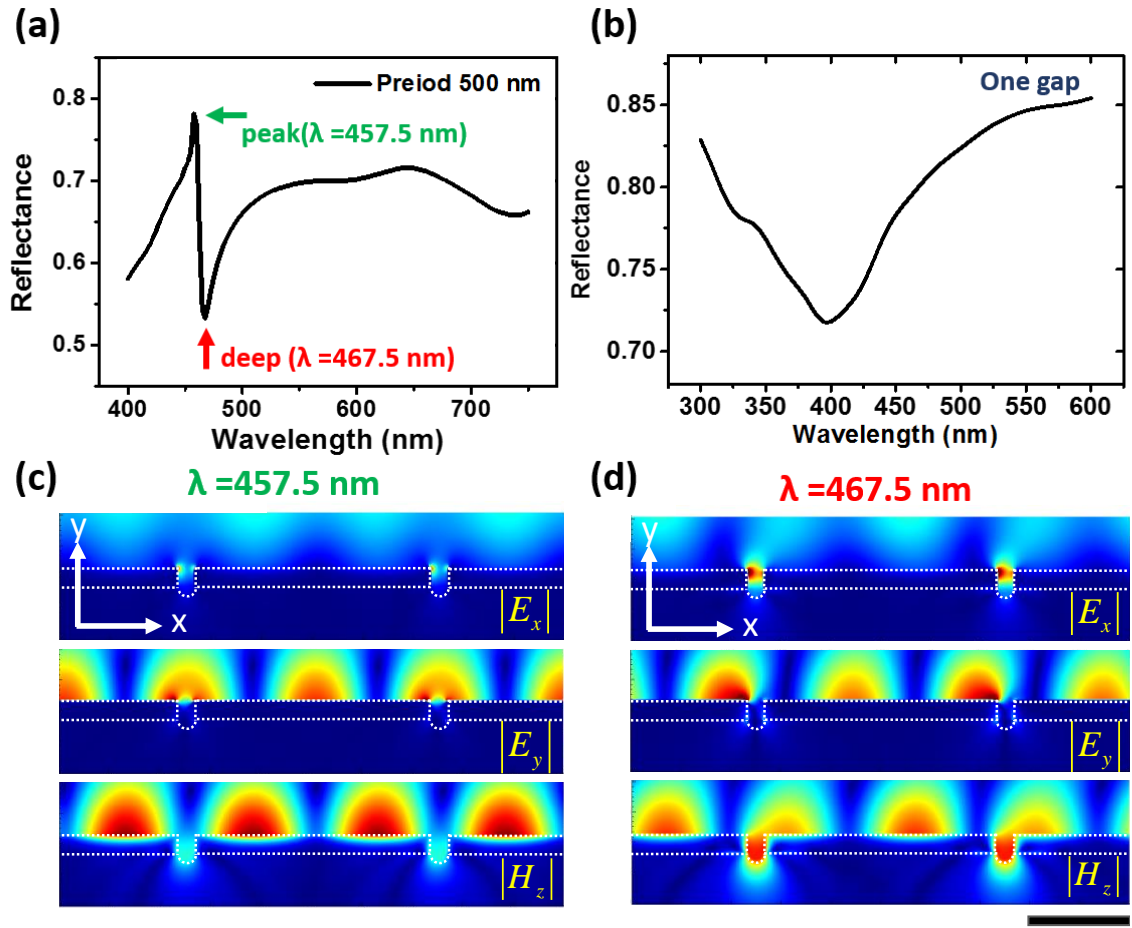

**Figure S4.** (a) The simulated spectrum of the reflectance from a nanoslits array with a periodicity of 500 nm on the aluminum film. (b) The simulated spectrum of the reflectance from a single nanoslit on the aluminum film. (c) and (d) The near-field distribution of $|E_x|$, $|E_y|$ and $|E_z|$ at the reflectance peak (λ=457.5 nm) and deep (λ=467.5 nm) in (a). Near-field $|E_x|$, $|E_y|$ and $|E_z|$ for the peak and dip are shown in the top, middle and bottom row in (c) and (d), respectively. Scale bar, 200 nm.

**(b) Fabry-Perot photonic modes and in-plane LSPR resonance of nanoholes**

To understand the origin of the peaks observed in Fig. 3(c) in the main text, we have performed simulations for a nanohole on a bare GaAs substrate without the aluminum film. Figure S5(a) depicts the schematic diagram of a nanohole structure used in the simulations. The nanohole (diameter = 300 nm, depth = 500 nm) is fabricated into the GaAs without and with the coverage of a 40-nm thick aluminum film. The nominal depth is defined as the distance from top surface to the bottom. Figure S5(b) shows the simulated scattering spectra of the nanohole fabricated in GaAs substrate with (black trace) and without (red trace) the aluminum layer. With aluminum layer, two peaks are observed around 430 nm and 520 nm. However, without aluminum film, only the peak around 430 is observed, suggesting that this mode is not due to LSPR of the aluminum hole. Figure 5S(c) shows the electric near-field intensity distribution at the two resonant wavelengths. Field distributions recorded at an x-y plane 5 nm away from the aluminum or GaAs surface are shown on the upper row and the cross sectional field distributions recorded with an x-z plane cutting through the nanohole center are shown in the lower row in Fig. 5S(c). From the near-field distribution, the broad peak around 520 nm (red dotted line) can be assigned exclusively to the LSPR and the sharp peak around 430 nm is due to the Fabry-Perot resonance of the photonic modes in the vertical air channel in the GaAs substrate. As can be seen, the photonic mode in the vertical air channel in GaAs is independent of the existence of the aluminum film. However, the sharp peak of photonic mode around 430 nm does shift to the red slightly with increasing the nanohole diameter, as can be seen in the Fig. 3(b) and 3(c) in the main text. This is because the effective index of the photonic mode increases slightly with increasing the diameter. The effective index of the photonic mode in the vertical air channel in GaAs with various diameters is summarized in Table S1. Note that the index is smaller than one because GaAs is an absorbing material.

**Table S1.** Effective index of the photonic mode in the vertical air channel in GaAs at various diameters.

| Diameter (nm) | 300 | 350 | 400 | 450 | 500 | 550 | 600 | 650 | 700 | 750 |
|---|---|---|---|---|---|---|---|---|---|---|
| Effective Index, n | 0.724 | 0.764 | 0.806 | 0.838 | 0.860 | 0.874 | 0.892 | 0.906 | 0.917 | 0.926 |

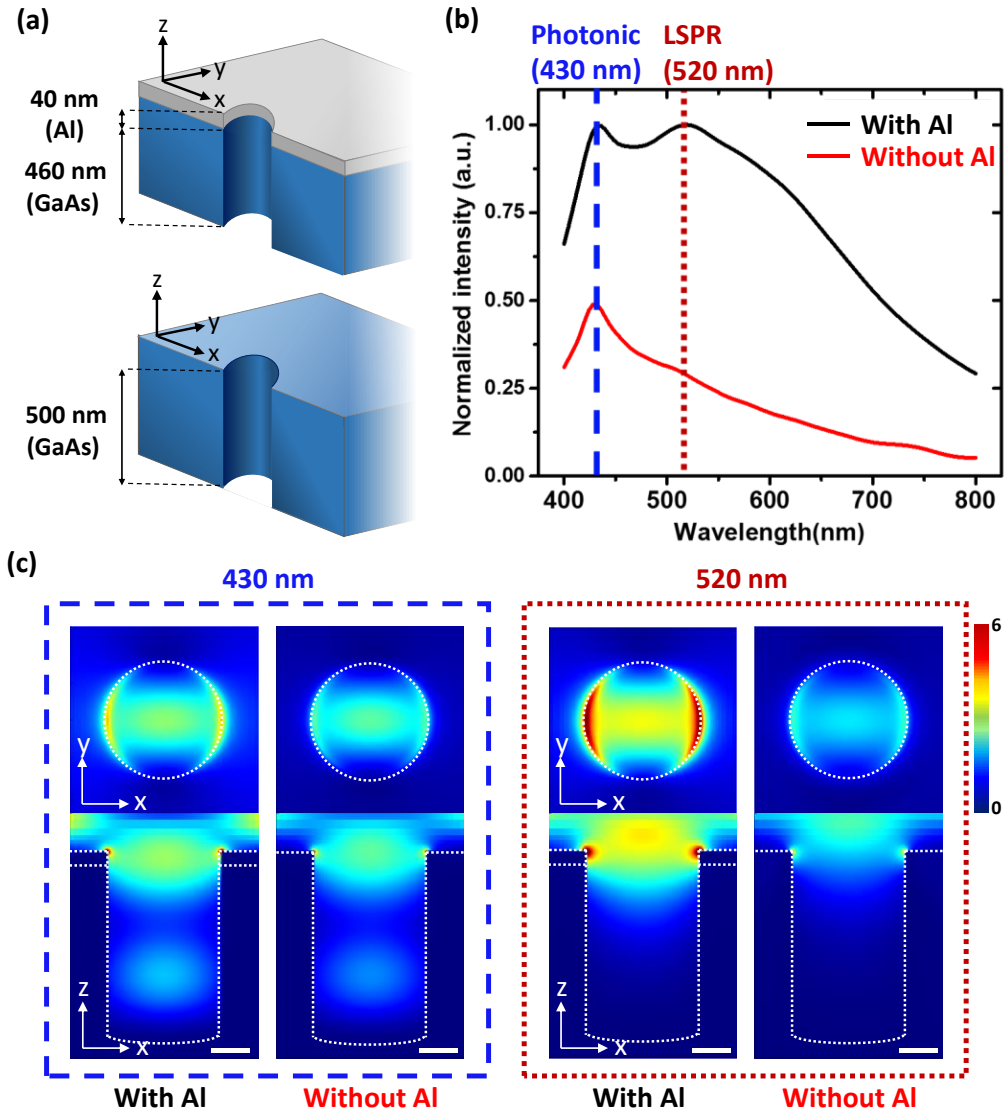

**Figure S5.** (a) The schematics of the nanohole (diameter = 300 nm) used in the FDTD simulations. The nanohole is fabricated into the GaAs with (top) and without (bottom) the coverage of a 40-nm thick aluminum film. (b) Simulated scattering spectra of the nanohole (diameter =300 nm, depth = 500 nm)fabricated into the GaAs substrate with (black sloid line) and without (red sloid line) the aluminum layer. The resonance of the photonic mode in the vertical air channel (blue dashed line) and the localized surface plasmon resonance (red dotted line) are denoted as "Photonic" and "LSPR", respectively. (c) The electric near- field intensity distribution (I = $|E_x|^2 + |E_y|^2 + |E_z|^2$ ) of the nanohole recored at the resonant wavelength of the photonic mode (430 nm, left) and the LSPR mode (520 nm, right). The upper row shows the intensity distribution recoreded at the x-y plane 5 nm away from the aluminum and GaAs surface. The lower row shows the intensity distribution recoreded in an x-z plane cutting through the center of the vertical air channel. Scale bar, 100 nm.

*TPPL spectrum*

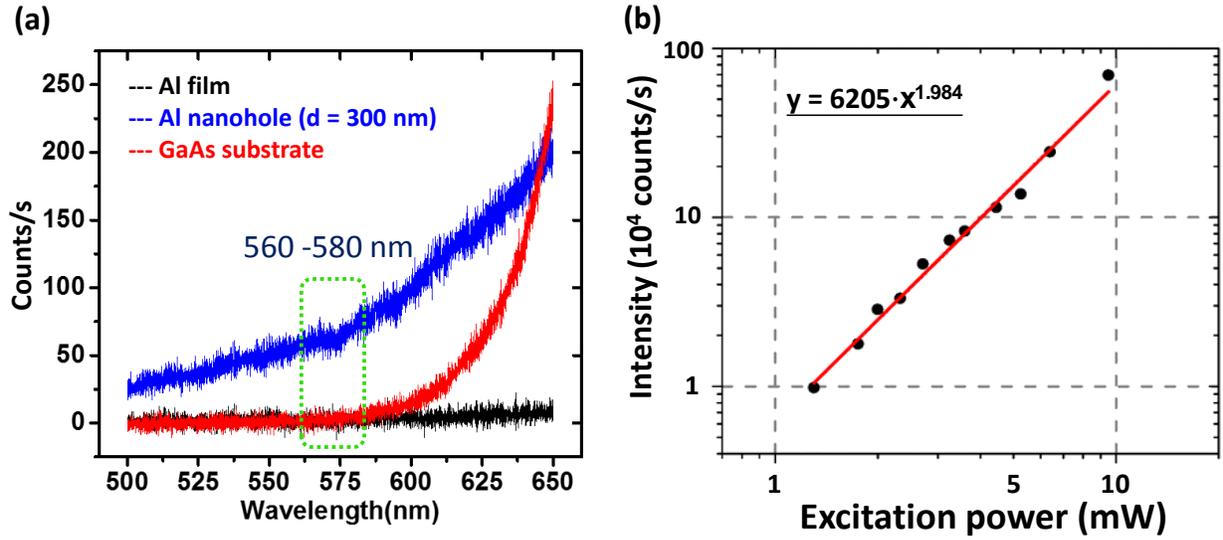

**Figure S6.** (a) The spectra of TPPL (excitation = 800 nm) from a flat area of single-crystalline aluminum film (black) and a nanohole with diameter of 300 nm on the single-crystalline aluminum film (blue), as well as that from a bare GaAs substrate (red). The intensity of the luminescence from the GaAs substrate increases rapidly around 580 nm and severely overlaps with the TPPL from aluminum nanostructures in the spectral region above 600 nm. To avoid the interference from the luminescence of GaAs substrate, the TPPL is collected by integrating only the spectrum with the window from 570 to 590 nm (marked by the green-dash rectangle). (b) The power dependence measurement of aluminum TPPL from wavelength 570 to 590 nm shows quadratic dependence.

*Mode analysis*

For small nanoholes, the relative large laser spot can most efficiently excite the dipolar resonant mode regardless of the excitation position, as shown in Fig. S7(a). However, for large nanoholes with diameter larger than the focal spot, the displacement of the laser spot from the geometrical center of the holes can lead to symmetry breaking of the system. This allow the excitation of the originally dark quadrupolar mode of the nanoholes, as shown in the Fig S7(b).

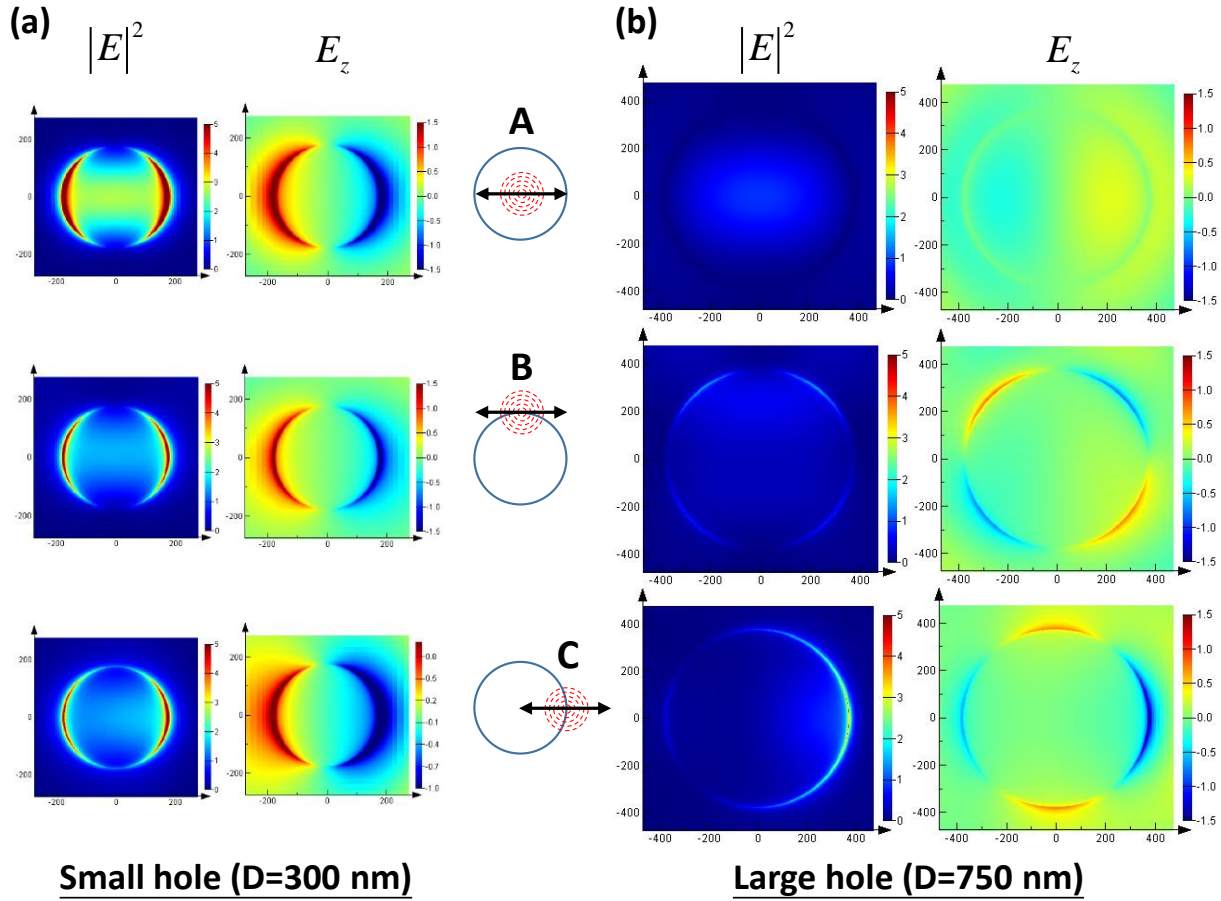

**Figure S7.** (a) The field distributions of a nanohole with diameter of 300 nm under the excitation of a tightly focused laser spot (beam diameter about 520 nm) at different excitation position. The near-field distribution shows a dipolar pattern. (b) The field distributions of a nanohole with diameter of 750 nm under the excitation of a tightly focused laser spot (beam diameter about 520 nm) at different excitation position. As long as the laser focal spot is displaced from the geometrical center of the hole (position B and C), the symmetry is broken and the dark quadrupolar mode can be excited. The near-field distribution shows a quadrupolar pattern.

*TPPL mapping setup*

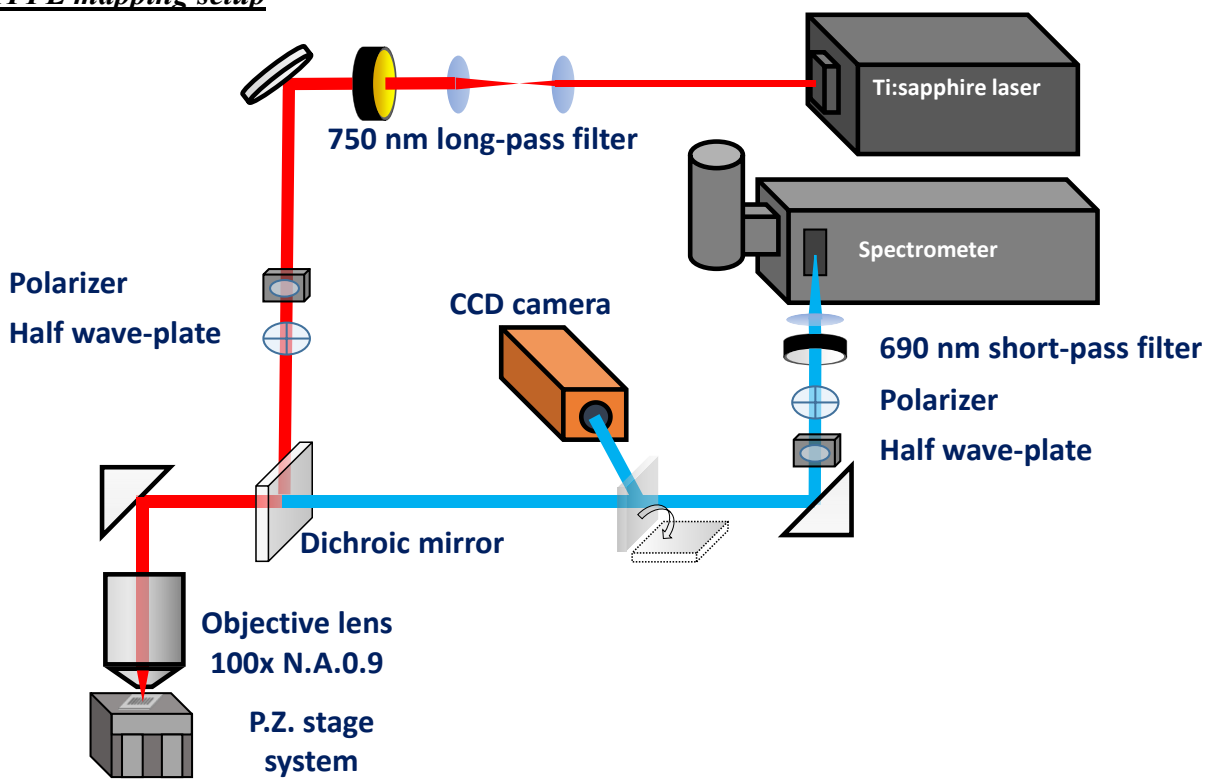

**Figure S8.** Confocal microscope setup for two-photon photoluminescence mapping.

*FDTD Settings*

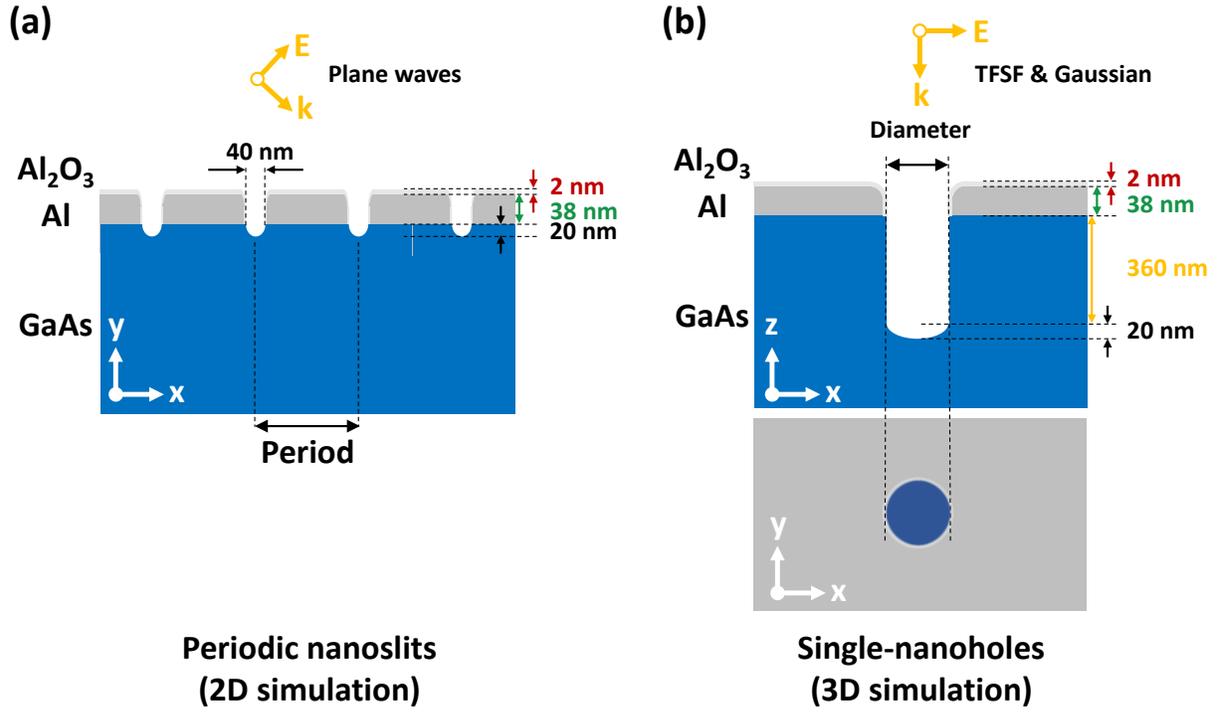

**Figure S9.** The excitation geometry used in the FDTD simulations for (a) the periodic nanoslits and (b) the nanoholes.